\documentclass[12pt,preprint]{aastex}

\usepackage{color}
\usepackage{amsmath}
\usepackage{float}
\usepackage{natbib}
\usepackage{ulem}
\usepackage{enumerate}

\def\me{$\,{\rm M}_{\oplus}\,$}
\def\re{$\,{\rm R}_{\oplus}\,$}
\def\sio2{SiO$_2$}
\def\gc3{g\,cm$^{-3}$}

\usepackage{xspace}
\def\sn{sub-Neptune\xspace}
\def\sns{sub-Neptunes\xspace}
\def\ssn{Sub-Neptune\xspace}

\newcommand{\ie}{i.e.,\xspace}

\newcommand{\eg}{e.g.,\xspace}

\begin{document}

\title{Contribution of the core to the thermal evolution of sub-Neptunes}
\author{A.~Vazan\thanks{email: a.vazan@uva.nl}, 
C.~W.~Ormel$^{1}$, L.~Noack$^{2}$, C.~Dominik$^{1}$\\
$^{1}$Astronomical Institute Anton Pannekoek, University of Amsterdam, The Netherlands\\
$^{2}$Department of Earth Sciences, Institute of Geological Sciences, FU Berlin, Germany}


\begin{abstract}
\ssn planets are a very common type of planets. They are inferred to harbour a primordial (H/He) envelope, on top of a (rocky) core, which dominates the mass.
Here, we investigate the long-term consequences of the core properties on the planet mass-radius relation.
We consider {the role of} various core energy sources {resulting} from core formation, its differentiation, its solidification (latent heat), core contraction and radioactive decay.
We divide the evolution of the rocky core into three phases: the formation phase, which sets the initial conditions, the magma ocean phase, characterized by rapid heat transport, and the solid state phase, where cooling is inefficient.
We find that for typical \sn planets of $\sim$2-10\me and envelope mass fractions of 0.5-10\% the magma ocean phase lasts several Gyrs, much longer than for terrestrial planets. 
The magma ocean phase effectively erases any signs of the initial core thermodynamic state. After solidification, the reduced heat flux from the rocky core causes a significant drop in the rocky core surface temperature, but its effect on the planet radius is limited.
In the long run, radioactive heating is the most significant core energy source in our model.
Overall, the long term radius uncertainty by core thermal effects is up to 15\%.
\end{abstract}

\section{Introduction}
Exoplanets in the range of several Earth masses are very common in our galaxy \citep{howard12,batlaha13,coughlin16}.
Observed mass-radius relation for some of these planets suggest that some of these planets are bare rocky planets with no envelope, and some are inferred to contain some amount of hydrogen and helium (\sns) on top of rock/iron core (hereafter {\it core}). The inferred envelope masses for \sns are typically of several percent \citep{lopez12,jontof16,wolfglopez15}, \ie most of the planet mass is in the core. In such a planet the core can perform as an energy  reservoir for the envelope.

In contrast to envelope (gases) cooling and contraction, the core radius is not expected to change much during the planetary evolution \citep{rogersseag10}. 
However, the heat flux from the core can affect the thermal properties of the low-mass envelope by heating it from below.
High enough heat flux from the core can lead to envelope mass loss \citep{ginzburg16,ginzburg17}, but even a moderate flux from the core can change the envelope thermal properties.
Hence, the thermal evolution of the core indirectly affect the planet radius.

Previous astrophysical studies have accounted for some of the core properties, like the decay of radioactive materials, when modeling the thermal evolution of the envelope \citep[\eg][]{lopezfor14,howebur15,chenrog16}, assuming an isothermal core that cools in the cooling rate of the envelope. 
However, in the mass range of Neptune-like planets, the core thermal properties can be significant \citep{baraffe08}.
A key factor in the core-envelope thermal evolution is thus the timescale on which the core releases its heat; if the cooling is on Gyrs timescale, the observed radius is affected by it \citep{vazan18a}. In order to study the timescale for the core cooling, its thermal evolution should be modeled explicitly.

In geophysical studies core heat transport is explicitly modelled \citep{turcotte67,stevenson83}. The cooling rate is determined by the properties of high viscosity convection and the resulting conductive boundary layers between convective regions. 
In these models the surface temperature of the planet is fixed as there is no thick envelope. In contrast,  in the case of \sn planets the surface temperature changes in time due to the envelope's cooling and contraction. 
The thick, gaseous envelope keeps the temperature (and the pressure) at the core surface higher than in an Earth-like case, affecting the viscosity and the state of the core (solid/liquid). Thus, for \sns the heat transport properties of the core depends on the envelope properties.
Therefore, both envelope and core should be modeled simultaneously.

In this work we model the thermal evolution of the planet as a whole, center to surface, and study the effects of the underlying thermal parameters on the state of the core, and on the derived planet radius.
In section~\ref{mdl} we discuss our model, and define different phases of the core thermal evolution.
In section~\ref{rslt} we show our results for \sn planets of $\sim$\,2\,-\,10\me with envelope mass fractions of 0.1\%\,-\,20\%, and examine thermal properties that lie within the uncertainty of geophysical models.
We discuss the \sns thermal evolution perspective in section~\ref{dscs}, and draw our conclusions in section~\ref{cncld}. 

\section{The model}\label{mdl}
\begin{figure}[ht]
\centerline{\includegraphics[width=12cm]{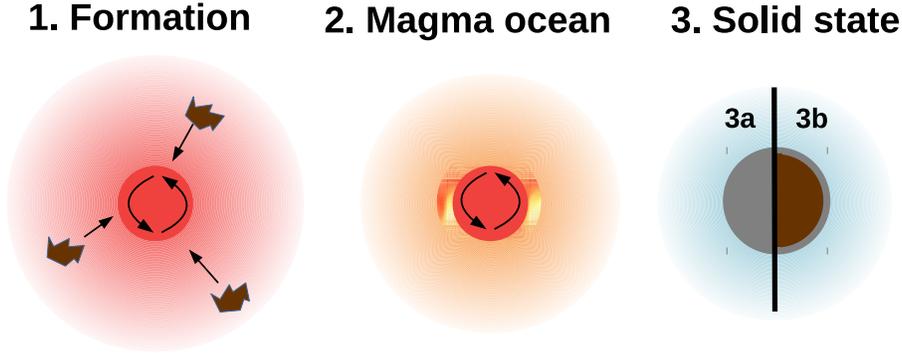}}
\caption{The three phases of the core evolution: 1. Formation - the core is hot due to conversion of binding energy to heat; 2. Magma ocean phases - vigorous (liquid) core convection; 3. Solid state - we consider either an entirely conductive core (3a), or a convective core with conductive core surface (3b).}
\label{fig_crtn}
\end{figure}

We calculate the thermal evolution of \sn planets with a rocky core and a hydrogen-helium envelope on a single structure grid (see Section~\ref{thrm_ev}).
As illustrated in Fig.~\ref{fig_crtn}, we divide the evolution of the planet into three phases:
\begin{enumerate}
\item Formation -- the conditions of the core and the gaseous envelope as derived from estimates of core formation (see Section~\ref{init}).
\item Magma ocean -- vigorous (liquid) core convection and efficient heat transport from the core to the gaseous envelope.
\item Solid state -- solidification of the core surface and conductive (reduced) heat transfer from the core to the envelope.
\end{enumerate}
The formation phase, in which the planet assembles in the presence of a gas-rich disk, is not explicitly modeled, but provides the initial conditions for the subsequent disk-free phases. The transition from phase 2 to phase 3 occurs when the core surface temperature drops below the solidification temperature for the surface pressure (see Section~\ref{flx_core}). This transition is approximated as instantaneous. The timing of the transition depends on the cooling of the core {\it and} the envelope, and on the core energy sources (see Section~\ref{enrg_core}). 
After solidification (phase 3) {core heat transport slows down significantly}. We distinguish two end-member scenarios: (3a) conductive core; (3b) convective core with conductive surface boundary layer.

In contrast to previous works, we determine the core thermal evolution from the perspective of the core thermal properties, and not from the envelope evolution. The change in core properties in time is simulated by the three phases.
The evolution path is continuous between the phases: at the end of each phase the structural parameters (radius, temperature, density, luminosity and composition for each mass layer) are being used for the first step of the next phase.

\subsection{Initial conditions}\label{init}
The initial energy content of the planet is determined by its formation. 
During the formation of a core the gravitational energy from the accumulation of the solid materials partially transforms to thermal energy. 
The speed of the accumulation process, as well as the material properties, determine which fraction of the gravitational binding energy is locked up in the planet in the form of thermal energy. 
A fast-forming core may retain large fraction of the initial binding energy and reach high initial temperatures. On the other hand, for a slow core formation a substantial amount of the binding energy is released already during the formation process (by radiation), which results in lower initial temperatures. 

The maximal temperature for the core formation can be estimated from the gravitational binding energy: 
\begin{equation}
E_{\rm binding}=\frac{3 G {M_c}^2}{5 R_c}
\end{equation}
The maximal temperature is achieved when all the binding energy is converted to heat, \ie $E_{\rm binding}$=$C_p M_c T_{\rm max}$ and thus  
\begin{equation}\label{eq_tmax}
    T_\mathrm{max} \sim\frac{G{M_c}}{R_c C_p} \\
    = 4.8\times10^4\,\mathrm{K}\ \left( \frac{M_c}{1\,M_\oplus} \right) \left( \frac{R_c}{1\,R_\oplus} \right)^{-1} \left( \frac{C_p}{1\,\mathrm{kJ\,(kg\,K)^{-1}}} \right)^{-1} 
\end{equation}
where M$_c$ and R$_c$ are the core mass and radius, $G$ is the gravitational constant, and C$_p$ is rock heat capacity \citep{guillot95}.
For a 4.5\me rocky planet of 1.65\re we get $T_\mathrm{max}=1.3\times10^5\,K$. 
This is rather high, since it assumes no radiative or advective losses.
But even when the initial temperature will be a fraction of $T_\mathrm{max}$, substantial amount of heat is locked in the core, and the rocky part will radiate at a high luminosity. 
Since the fraction of gravitational binding energy that is left in the core after formation is unknown, we consider here a wide range of fractions between 5\%-50\% of $E_{\rm binding}$.

The core is embedded in an initially diluted adiabatic envelope. 
We assume that the planet envelope is formed by gas accretion during the disk phase \citep{pollack96,ikoma12}, and therefore is composed of hydrogen and helium at solar ratio. The envelope extends initially to the planet Hill radius\footnote{We calculated also for initial radius of 0.1 R$_{\rm Hill}$, and find very minor effect on the long-term evolution for all cases.}. 

\subsection{Flux from the core}\label{flx_core}
Heat transport by convection in the core is determined by the Rayleigh number 
\begin{equation}\label{eq_Ra}
{\rm Ra}=\frac{{\rho}{\alpha}g{\Delta}TD^3}{\kappa\eta},
\end{equation}
which depends on structure properties as gravity ($g$), density ($\rho$), layer thickness ($D$) and the temperature difference within the layer  (${\Delta}T$), but also on material properties such as thermal expansion coefficient ($\alpha$), thermal diffusivity ($\kappa$) and kinematic viscosity ($\eta$).  
Convection occurs for Rayleigh numbers greater than a critical value (here we use $Ra_{\rm crit}=450$, \cite{morschhauser11}). 
According to Eq.~\ref{eq_Ra} convection is less likely and less vigorous if viscosity $\eta$ and thermal conductivity $\kappa$ are high.
Between these two, the viscosity is the key factor in regulating the convection, since it is an exponential function of the local physical conditions.
Specifically, we approximated the viscosity by \citep[\eg][]{karatowu93,poirier00,noack16}
\footnote{In high pressures as in the inner part of \sns, viscosity also strongly depends on pressure \citep{stamenkov11,tackley13}. Here we neglect the pressure term in the viscosity equation, because we focus on the conditions at the core-envelope-boundary, where the pressure is not more than several GPa for \sn mass range. This may slightly overestimate the cooling of the core.}:
\begin{equation}\label{eq_vsc}
\eta(T)=\eta_0 \,\, \exp\left[\frac{E_0}{R_g}\left(\frac{1}{T}-\frac{1}{T_0}\right)\right],
\end{equation}
{where} reference values are extrapolated from Earth-like composition: viscosity $\eta_0=10^{21}\,{\rm Pa\,s}$, temperature $T_0=1600\,$K, and activation energy $E_0=240\,$kJ/mol; $R_g$ is the gas constant.

In the core-envelope-boundary (hereafter CEB) heat is transported by conduction. The thickness of the conductive layer $\delta$ is determined by the vigor of the underlying convection \citep{stevenson83}:
 \begin{equation}\label{eq_dlt}
\delta={D}\left(\frac{Ra_{\rm crit}}{Ra}\right)^{1/3}
\end{equation}  
As the core cools and the viscosity increases, the Rayleigh number is lower and thus the conductive boundary layer becomes thicker.
Since heat transport in the boundary layer operates by conduction -- a diffusive process -- core cooling depends on the conductive timescale, which is in the order of $\tau_{\rm cond}={\delta^2}/{\kappa}$. 

Initially, the core is in a {\it magma ocean} phase -- a fluid phase of very low viscosity -- due to its high temperature from formation. In this phase the conductive boundary layer is very thin, and thus the conductive timescale is short ($\tau_{\rm cond}\leqslant10^6\,yr$).
In this phase we assume efficient heat transport from the core to the envelope, and in the core, \ie we simply ignore the conductive CEB and model the core as adiabatic.

{The magma ocean phase continues until at some point the core is cold enough to allow solidification of the core surface.
To find this point we compare the CEB temperature with the critical melting temperature for Earth mantle composition, as described in Appendix A1.
The magma ocean solidifies from the bottom upward, since the melting temperature of rock increases more strongly with pressure than the increase in temperature of the adiabatic core structure.
Therefore, CEB solidification implies that the core can no longer be modeled by low viscosity convection. 

After solidification, the high viscosity of the rock slows down the heat transport in the core. The change in the core heat transport after solidification is a gradual process, which depends on uncertain structure and material properties\footnote{Solidification also depends on the rock exact composition; very different composition can lead to different solidification temperatures and thus earlier/later solid state.}.
For simplicity, we consider here two extreme limits:\\
{\bf(3a)} conductive core, where we assume that convection will be suppressed due to the high interior pressure \citep[\eg][]{vandenBerg10,stamenkovic12}.\\
{\bf(3b)} convecting core, with conductive thermal CEB layer (including the crust) on top of it.\\
We model the transition between the magma ocean phase (phase 2) and the solid state phase (phase 3) as instantaneous.

In the case of an entirely conductive core we take the heat transport in all core layers to be conductive. We set the actual temperature gradient (Eq.~\ref{eq_tgrad}-\ref{eq_nablaR} in Section~\ref{thrm_ev} below) to conductive heat transport, using Earth mantle conductivity (Table~\ref{tab_mdl}).
{In the case of conductive CEB layer} we assume a conductive CEB layer on top of a convecting core. The thickness of the conductive layer is calculated according to Eq.~\ref{eq_dlt} at the CEB solidification point. We take the heat transport only in that layer to be conductive, while the rest of the core remains convective.

\begin{table}[t]
\centering 
\begin{tabular}{l c c} 
 \hline\hline\\[-1.5ex] 
Sources & Energy & Release timescale \\         
& [J/kg] & [yr] \\
\hline\\[-1.5ex]
    Formation$^{a}$     & $3.4\times 10^7$  &  $0$ \\
    Differentiation     & $3.1\times 10^6$  & $0 - 10^7$ \\
    Radioactive decay$^b$ & $2.6\times 10^6$  & $10^9$   \\
    Solidification      & $6.0\times 10^5$  &  at $t_\mathrm{solid}$ \\
    Contraction$^{c}$   & $5.5\times 10^4$  &  self-consistent  \\
 \hline 
  \\[0.01ex] 
\end{tabular}
\caption{Core energy sources}
\tablecomments{Values are for a 5\me planet with 10\% envelope.\\ $^a$ Core accretion energy, value for $E_{\rm acc}$=$0.2\,E_{\rm binding}$.\\ $^b$ Earth-like abundance. \\ $^c$ Automatically incorporated in model. Energy contribution is estimated.}
\label{tab_Es}
\end{table}

\subsection{Core energy sources}\label{enrg_core}
 
Our model includes several energy sources. The magnitude of the involved energies and the timescale on which they operate affect the state of the core and the envelope. Below are the sources included in our model:

\begin{enumerate}[I]
\item {\bf Formation}
As discussed in Section~\ref{init}, the exact fraction of the accretion energy to be locked in the core depends on formation and thus unknown. 
Therefore, we calculate evolutionary tracks for planets with different fractions of the binding energy from core formation, in the range of 0.05-0.5\,$E_{\rm binding}$. 
The lower bound of this range is calculated from the minimal energy of early Earth-like geophysical models \citep{noack17}, and the upper bound is an overestimation, assuming half of the impact energy is left in the core. We calculate models for values of 0.05, 0.1 0.2 0.3, 0.5\,$E_{\rm binding}$.

\item {\bf Differentiation}
The core is initially of low viscosity (molten). Under this condition iron sinks to the center of the planet very efficiently \citep{stevenson90}, \ie during the early evolution.
The released gravitational energy further heats the interior. We use the formalism of \cite{solomon79} to add the differentiation energy to the core. We assume the iron-to-rock ratio to be Earth-like, and find the differentiation energy to be of a few percent of the binding energy.
For simplicity, we add this energy at once {(\ie at $t=0$)}, to the initial energy content of the core\footnote{Since the exact time of differentiation ranges between 10$^6$-10$^7$\,yr (ref), we calculated also for later deposition of the differentiation energy, at t=10$^7$\,yr, and find the long term (Gyrs) evolution to be the same.}.
 
\item {\bf Radioactive heating}
The decay of radioactive elements is an important heat source in rocky planetary interiors \citep{valencia07c}. 
The dominant elements with half lives in the Gyrs regime are  $\rm ^{\rm 238}U,\,^{\rm 235}U,\,^{\rm 232}Th$ and $^{\rm 40}K$ \citep{nettel11,andgrev89}. 
We apply the radiogenic luminosity to Eq.~\ref{eq_Lc} (see below), by using the values of \cite{nettel11}.
The abundances of radioactive elements for exoplanets are unknown, but can range between 0.5 and 2.5 times the Earth ratio in solar analogue stars \citep{unterborn15}.
Therefore, we calculate for planets in this abundance range. 

\item {\bf Solidification (latent heat)}
As the planet cools the core changes from liquid to solid. The solidification process releases latent heat. We include the latent heat release in our model by adding it to each planetary layer that cools below the solidification temperature. We use the melting curve as provided in Appendix A1.
Rock latent heat of $6\times10^5\,J/{\rm kg}$ \citep{morschhauser11} is added to the luminosity (Eq.~\ref{eq_Lc}) of each solidified layer.

\item {\bf Core contraction} 
The pressures in \sn interiors can reach GPa levels, where core compression may heat the core and affect the thermal evolution \citep{mordasini12}.
In our calculation this effect is automatically included since the heavy element core is part of the evolution structure matrix, and pressure-temperature-density relation are derived from the rock EOS \citep{vazan13}. 
Estimate of the energy from core contraction (by $p\,\Delta V$) is found to be on the order of $10^{4}-10^{5}\,J/{\rm kg}$ for \sn mass range.
\end{enumerate}

Estimates of the above energy sources and their release timescales,  for a 5\me planet with 10\% envelope, appear in Table~\ref{tab_Es}. 
The fits for the energy flux are implemented in the model by using the method described in \cite{vazan18a}. 
Collecting these effects, the luminosity in the core is taken to be:
\begin{equation} \label{eq_Lc}
L_\mathrm{core} = M_c \,\left(c_v \frac{dT_\mathrm{c}}{dt}+ \frac{E_{\rm radio}}{\tau_{r}} e^{\left(-t / \tau_{r} \right)} +\frac{E_{\rm solid}}{\Delta t }\,\delta\left(T-T_{\rm solid}\right)\right)
\end{equation} 
where $c_v$ is the specific heat capacity, $M_c$ the core mass, and $t$ is time.
$dT_{c}/dt$ describes the release of initial energy from formation and differentiation, 
$E_{\rm radio}$ and $\tau_{r}$ are adjusted to fit the heat production by radioactive decay as in \cite{nettel11}, and $E_{\rm solid}$ is the solidification (latent) energy \citep{morschhauser11} released on time interval $\Delta t$, when the temperature reaches the solidification temperature (T=T$_{\rm solid}$). The parameter values appear in Table~\ref{tab_mdl}.
External core energy sources, such as late planetesimal capture \citep{chattchen18}, are not included in the model.

\subsection{Thermal evolution}\label{thrm_ev}
The evolution is modelled by a 1D hydrostatic planetary code that solves the structure and evolution equations for the entire planet on one grid \citep[see][for details]{kovetz09,vazan13,vazan15}.
We use an equation of state (EOS) for hydrogen, helium \citep{scvh} and rock, as described in \cite{vazan13}. 
The thermodynamic properties of the core (such as density, entropy, etc.) are modeled by the EOS of one material (\sio2), as a simplification for the full mineralogy of a core.

The energy balance during the evolution is described by 
\begin{equation}\label{eq_ebalance}
\frac{\partial u}{\partial t}+p\frac{\partial}{\partial t} \frac{1}{\rho}=q-\frac{\partial L}{\partial m}
\end{equation}
where the symbols $\rho,p,u$ are density, pressure and specific energy, respectively, $q$ is the contribution by the additional energy sources, and $m, L$ are the planetary mass and luminosity.

The temperature profile is determined by the heat transport rate according to 
\begin{equation}\label{eq_tgrad}
\frac{\partial \ln T}{\partial m}=\nabla \frac{\partial \ln p}{\partial m}
\end{equation}
In convective regions the actual temperature gradient $\nabla$ is the adiabatic temperature gradient $\nabla_A$; otherwise, heat transfers via radiation (in the envelope) and conduction (in the core), i.e. $\nabla=\nabla_R$, where 
\begin{equation}
\nabla_R=\frac{\kappa_{\rm op} L}{4\pi cGm}\frac{p}{4p_R}.
\label{eq_nablaR}
\end{equation}
$\kappa_{\rm op}$ is the harmonic mean of radiative and conductive opacities, and $p_{\rm R}$ is the radiation pressure. 

\subsubsection{Atmospheric conditions}
{The} atmosphere opacity regulates the planet luminosity and the contraction of the envelope. We use Rosseland mean of the radiative opacity by \cite{sharp07} for solar metallicity grain-free atmosphere.
Envelopes of \sns may have an enhanced atmospheric metallicity from their formation \citep{fortney13,thorngren16}.
Moreover, extended clouds in \sn atmospheres \citep[\eg][]{bean11,desert11} indicate large atmospheric metallicities and a large opacity \citep{morley13}. 
Because of the important role of atmospheric opacity on the cooling rate of planets, we tested the sensitivity of our results to higher (30$\times$solar) atmospheric metallicity.

At the atmospheric boundary, which is taken to be the planetary photosphere, the outer boundary condition is 
$ \kappa_{\rm op} p = \tau_s g$ where $g=GM/R^2$ is the gravitational acceleration, and $\tau_s$ is the optical depth of the photosphere. 
The envelope mass is constant during the evolution, \ie no evaporation or gas accretion are included in the model. 

\subsubsection{Irradiation}
We assume the temperature distribution in a gray, plane-parallel atmosphere, with constant net outward flux $F$ and irradiation temperature $T_{\rm irr}$, to be 
\begin{equation}\label{eq_bc6}
\sigma T^4(\tau)=\left(\frac{3}{4}\tau+\frac{1}{4}\right)F+g(\tau)\sigma T_{\rm irr}^4.
\end{equation}
where $g(\tau)=\frac{3}{2}(1-\frac{1}{2} e^{-\tau})$ \citep{kovetz88} and $\sigma$ is Stefan-Boltzmann's constant.
The temperature distribution is calculated for a vertical (maximal) irradiation flux, with no angle dependency of the incident flux \citep{guillot10}.
At the photosphere, where $\tau=\tau_S=1\,$, the net outward luminosity of the planet is
\begin{equation}\label{eq_bc7}
L = 4\pi R^2F = 4\pi R^2\sigma \left[T^4-g(\tau_S) T_{\rm irr}^4\right].
\end{equation}

The irradiation temperature as a function of the distance from the star is
\begin{equation}\label{eq_tirrd}
T_{\rm irr}=\left(\frac{L_{\star}(1-A)}{16\pi\sigma d^2}\right)^{1/4}
\end{equation}
where $L_{\star}$ is the stellar luminosity (we use $L_{\star}=L_{\sun}$), $d$ is the distance of the planet from the star, and $A$ is the albedo. 
In order to separate the core effects from the environmental thermal effects, we first take d=1\,AU as our standard model. This corresponds to the outer edge of Kepler's detection region. Next, we test closer-in cases of d=0.3\,AU and examine the effect on the results. We avoid planets at d$<$0.3\,AU, from which photoevaporation can become significant,  
since photoevaporation by the parent star removes (part of) the gaseous envelope \citep{owenwu13,owen16}, and thus changes the envelope mass in time.

The albedo strongly depends on the atmospheric composition and is an unknown parameter for \sns. 
Here we take all the irradiation to be absorbed by the planet ($A=0$), which is the upper bound of the irradiation effect. 
Thus, a non-zero albedo planet should be located closer-in for an equivalent irradiation flux.

\begin{table}[t]
\centering 
\begin{tabular}{l c c c} 
 \hline\hline\\[-1.5ex] 
Parameter & value & unit & ref\\         
\hline\\[-1.5ex]
$E_{\rm radio}$ (Eq.~\ref{eq_Lc}) & 2.64$\times$10$^{10}$ & erg/g & 1\\
$\tau_{r}$ (Eq.~\ref{eq_Lc}) & 1.85$\times$10$^{9}$ & yr & 1\\
$E_{\rm solid}$ (Eq.~\ref{eq_Lc}) & 6$\times$10$^{9}$ & erg/g & 2\\
\hline\\[-1.5ex]
L$_{{\sun}}$ & 3.8515$\times10^{26}$ & W & 3\\ 
$\kappa_{\rm op}$ & Z$_{\sun}$ - 30$\times$Z$_{\sun}$ & cm$^2$/g & 4\\
k$_{\rm cond}$ & 4 & W/m/K & 5\\
d & 0.3 - 1 & AU\\
albedo & 0 & \\
 \hline 
  \\[0.01ex] 
\end{tabular}
\caption{Model parameters}
\tablecomments{Set of parameters we use in the model. References: $^1$\cite{nettel11}, $^2$\cite{morschhauser11}, $^3$\cite{guenther92}, $^4$\cite{sharp07}, $^5$\cite{stevenson83}. 
}\label{tab_mdl}
\end{table}

\section{Results}\label{rslt}

\subsection{Magma ocean phase}
\subsubsection{Role of core energy sources} 
In Fig.~\ref{fig_std} we show the evolution of 5\me planets with 90\% (4.5\me) core surrounded by 10\% (0.5\me) of hydrogen-helium envelope, located at 1\,AU from a Sun-like star. 
The radius (left) and core surface (CEB) temperature (right) are shown in the figure, where the different curves are for different approximations for core energy sources. 
The CEB solidification occurs when the CEB temperature crosses the dashed horizontal melting line. At this time point (which we call $t_\mathrm{MO}$) the magma ocean phase ends.
At first, all cases in the figure are modeled as in the magma ocean phase, \ie efficient convection{, even for $t>t_\mathrm{MO}$}.
Thus, the cooling after the solidification point (thin curves) is overestimated.
In Section~\ref{rsolid} below we show the effect of the solidification on the results. 

The lowest core energy case (blue curve) is achieved for the lower bound of 0.5\,Earth radioactive abundance and minimal initial energy from formation ($E_{\rm acc}$=$0.05 E_{\rm binding}$).
Then, we increase the initial energy content of the core to 20\% of the binding energy ($E_{\rm acc}$=$0.2 E_{\rm binding}$) and include the differentiation energy (red-dotted {curve}). As is shown in the figure, the long term evolution is the same as the previous case, since this energy is being released efficiently during the magma ocean phase on a timescale of $\sim10^7$yr. 
Although the initial core energy content doesn't affect the long term evolution, this initial energy can expand the envelope until it isn't bound to the planet. We test this idea (see appendix A2), and find that for an initial energy of more than 30\% of the binding energy, part of the envelope is being lost, for the planets in Fig~\ref{fig_std}.

\begin{figure}[ht]
\centerline{\includegraphics[width=12cm]{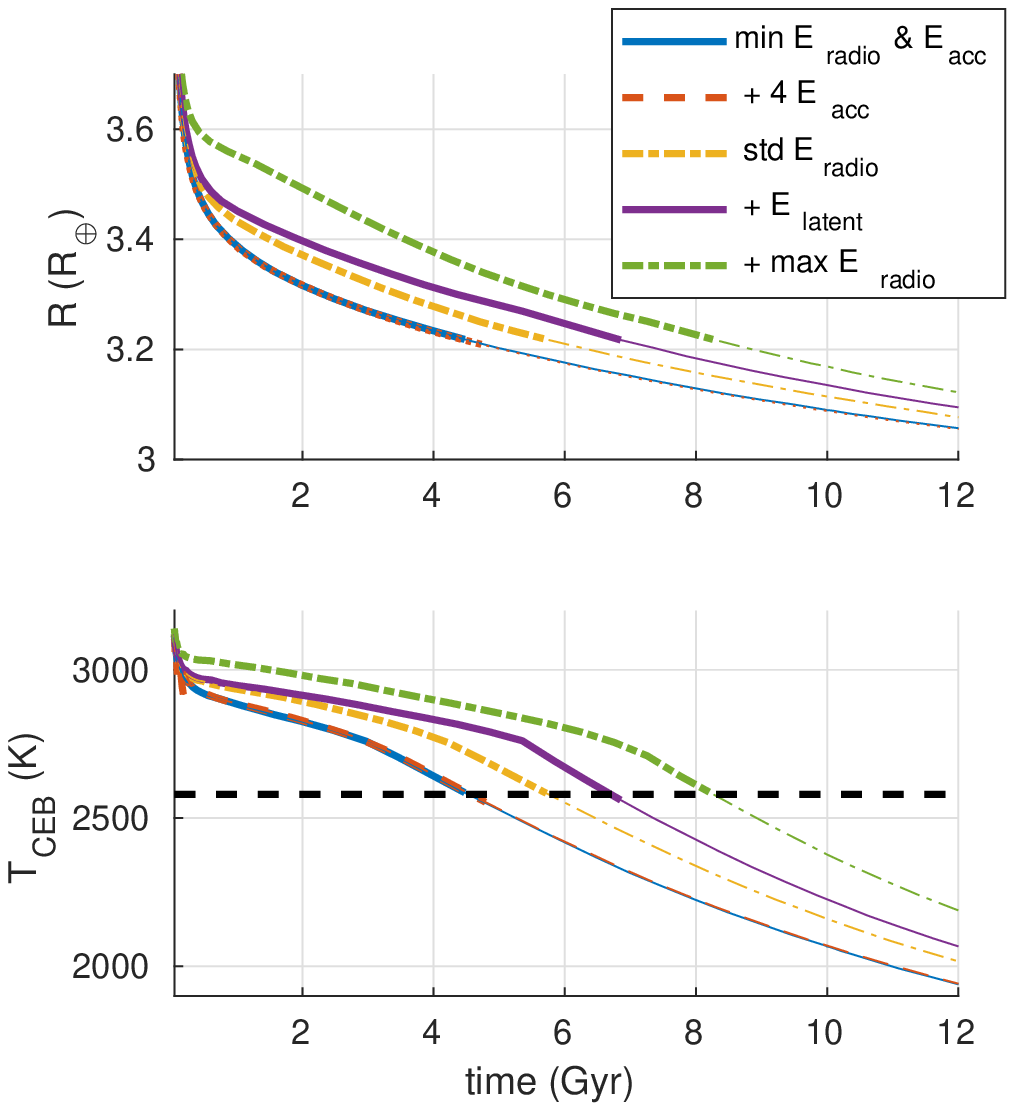}}
\caption{Radius (left) and CEB temperature (right) of 4.5\me cores with 0.5\me (10\%) hydrogen-helium envelope. 
The different curves are for different assumptions for core energy sources: minimal radioactive heating and low formation energy ($E_{\rm acc}$=$0.05 E_{\rm binding}$) (blue), 4 times higher accretion energy ($E_{\rm acc}$=$0.2 E_{\rm binding}$) and including differentiation energy (red-dashed), Earth ratio of radioactive heating (yellow-dashed), including latent heat (blue), and maximal radioactive heating together with all previous energy sources (green-dashed).
The horizontal line indicates the core surface solidification.} 
\label{fig_std}
\end{figure}

{Next}, we increase the {contribution from} radioactive element {decay} to {levels similar to} Earth (dashed-yellow). Since the radioactive energy is being released on Gyr timescale, the long term radius and CEB temperature are higher for this case.
In the next case (purple) we added the latent heat by core solidification. 
The latent heat release by core solidification occurs from inside-out.
This is a gradual process during the magma ocean phase, which ends when the CEB solidifies. 
The latent heat release delays somewhat the CEB solidification by up to 1\,Gyr, and slightly increases the radius.
This case (purple curve) of a core with Earth abundance of radioactive content, and a latent heat release at solidification is our standard model.

{Finally, we considered a model with radioactive levels enhanced by a factor 2.5 over Earth.}
This high radioactive level is motivated by the maximal measured abundance around sun-like stars \citep{unterborn15}.
{This enhanced} radioactive heating (dashed-green) in the core is found to be the most important energy contribution to the long term evolution.
Variation in the fraction of radioactive elements in the observed range of 0.5-2.5$\times$ Earth abundance, results in up to 5\% radius change for planets with solar-metallicity envelope. 
In general, fraction of radioactive elements and solidification (latent) heat are the most significant core energy sources for the long term. In addition to {inflating} the radius, the excess heat delays the CEB solidification time (from 4\,Gyr to 8\,Gyr in this case), and thus prolongs the magma ocean phase. {For these cases, many observed sub-Neptunes are likely to be in a magma ocean phase.}
 
\subsubsection{Dependence on the envelope mass}
In Fig.~\ref{fig_Tceb} we follow the change in surface (CEB) temperature during planet evolution, of planets with same core mass surrounded by envelopes of different masses. 
We use the 4.5\me standard core model (purple curve in Fig.~\ref{fig_std}), and vary the mass of the envelope between 0.1\%-20\% (0.005-1\me).
The horizontal curves in the figure show the point when the CEB solidifies, for the surface pressure and temperature of this planet, as described in appendix A1.
In these runs, as before, magma ocean phase conditions are assumed for the entire evolution, thus the cooling after solidification (thin curves) is overestimated.

As is shown in the figure, the magma ocean phase (thick part of each curve) for planets with significant envelopes {lasts} much longer than in the case of Earth. 
As the atmospheric mass increases, the temperature on the surface of the core is higher and, as a result, the surface stays molten for longer time. Therefore, for most of the cases in Fig.~\ref{fig_Tceb} the duration of the magma ocean phase increases with envelope mass.

\begin{figure}[ht]
\centerline{\includegraphics[width=12cm]{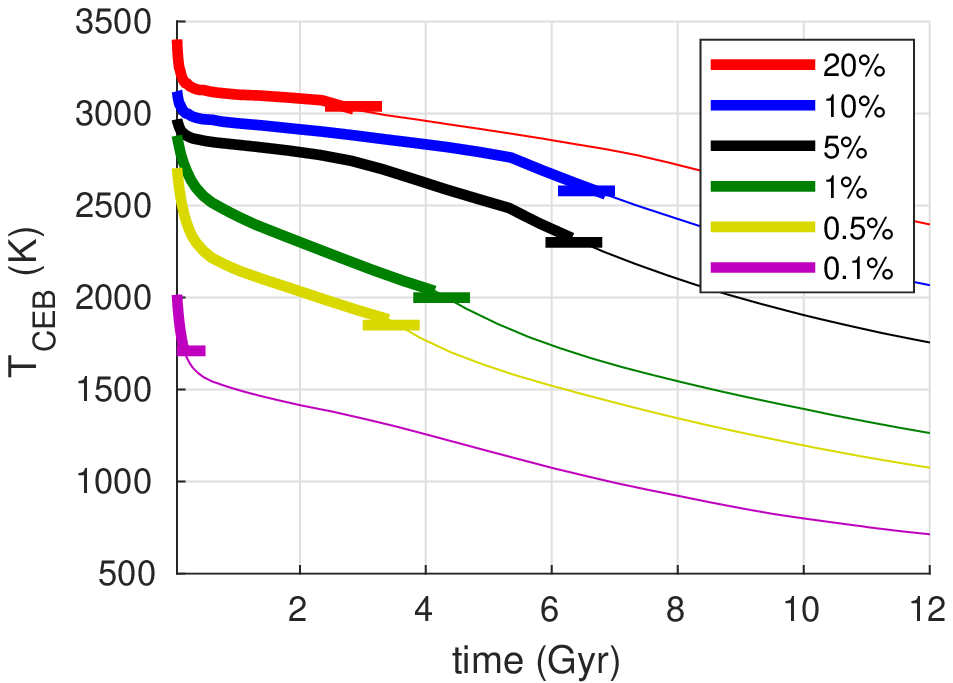}}
\caption{CEB temperature of 4.5\me cores with different envelope masses. The percentages denote the mass fractions of the envelope with respect to the total planet mass. The horizontal lines indicate the (pressure-dependent) CEB solidification temperature of each planet (see appendix A1). The evolution is modelled by efficient magma ocean cooling, which overestimates the cooling after solidification ({thin} curves).} 
\label{fig_Tceb}
\end{figure}

However, with {increasing} envelope mass, the pressure on the planet surface also increases and {so does} the melting temperature. 
For a thick envelope of about 1\me (red curve) the melting temperature is above 3000\,K and therefore the CEB solidifies {\it earlier} than for lower mass envelopes.  
As a result, the magma ocean duration is limited to up to several Gyrs, since for high mass envelopes ($>$0.5\me) the required higher melting temperatures shorten the magma ocean phase.
We find that \sn planets with envelope masses between 0.01\me$<M_{\rm env}<$1\me make the transition from magma ocean to solidified state during the time we observe them (1-7\,Gyr).
In the next section we estimate the effect of such a transition on the properties of the planet.

\subsection{Solid state phase}\label{rsolid}
In Fig.~\ref{fig_TR_d} we present the planetary radius (left) and the CEB temperature (right) for a 4.5\me core with 0.5\me (10\%) and 0.05\me (1\%) envelopes. The solid curves represent the efficient core cooling as in the magma ocean phase (same as the blue and green curves in Fig.~\ref{fig_Tceb}). 
After the CEB solidifies (dashed horizontal temperature line) we calculate for {the} two scenarios described in Section~\ref{flx_core}: thermal evolution with entirely conductive core (dashed-dotted), and thermal evolution with conductive CEB layer (dashed).
The thickness of the conductive layer, according to Eq.~\ref{eq_dlt}, is about 100\,km, and the timescale for cooling by conduction through this layer is on the order of 10$^8$\, yr.  

As is shown in the figure, the CEB temperature {rapidly} changes {at the point where the model (suddenly) assumes a conductive structure}. The conductive CEB {scenario} moderates the flux from the core but keeps the long term surface temperature similar to the magma ocean case. The conductive boundary layer acts as a bottle-neck for the cooling, but it is not thick enough to slow the cooling substantially.
In the conductive core scenario the cooling is much slower{;} heat is locked in the core while the surface temperature drops. 

The effect of solidification on the thermal evolution depends on the envelope mass.
In the case of the 0.5\me envelope ({upper panels}) the CEB solidification occurs {at a} later {time} than in the 0.05\me envelope case. At a later time the core energy {budget} is smaller (mainly {due to} radioactive decay). In addition, the cooling at the bottom of the thicker envelope (the core interface) is slower. Therefore, the drop in CEB temperature is more moderate. 

While the effect of the core properties is significant for some of the cases, the effect on the radius is {more} limited. The maximum radius {change} for \sn planets in our sample {as resulting from solidification} is about 6\% percent, {or} 0.1\re. 
{This} maximum is achieved for a planet with 0.05\me envelope mass (bottom left in Fig.~\ref{fig_TR_d}). 
{For planets with lower mass envelopes, the contribution of the envelope to the total radius is small. On the other hand, planets with} high envelopes masses, such as the 0.5\me envelope case (upper left in Fig.~\ref{fig_TR_d}), have lower core-to-envelope mass ratios and smaller energy contents at solidification, which limit the effect on the radius.

\begin{figure}[ht]
\centerline{\includegraphics[width=12cm]{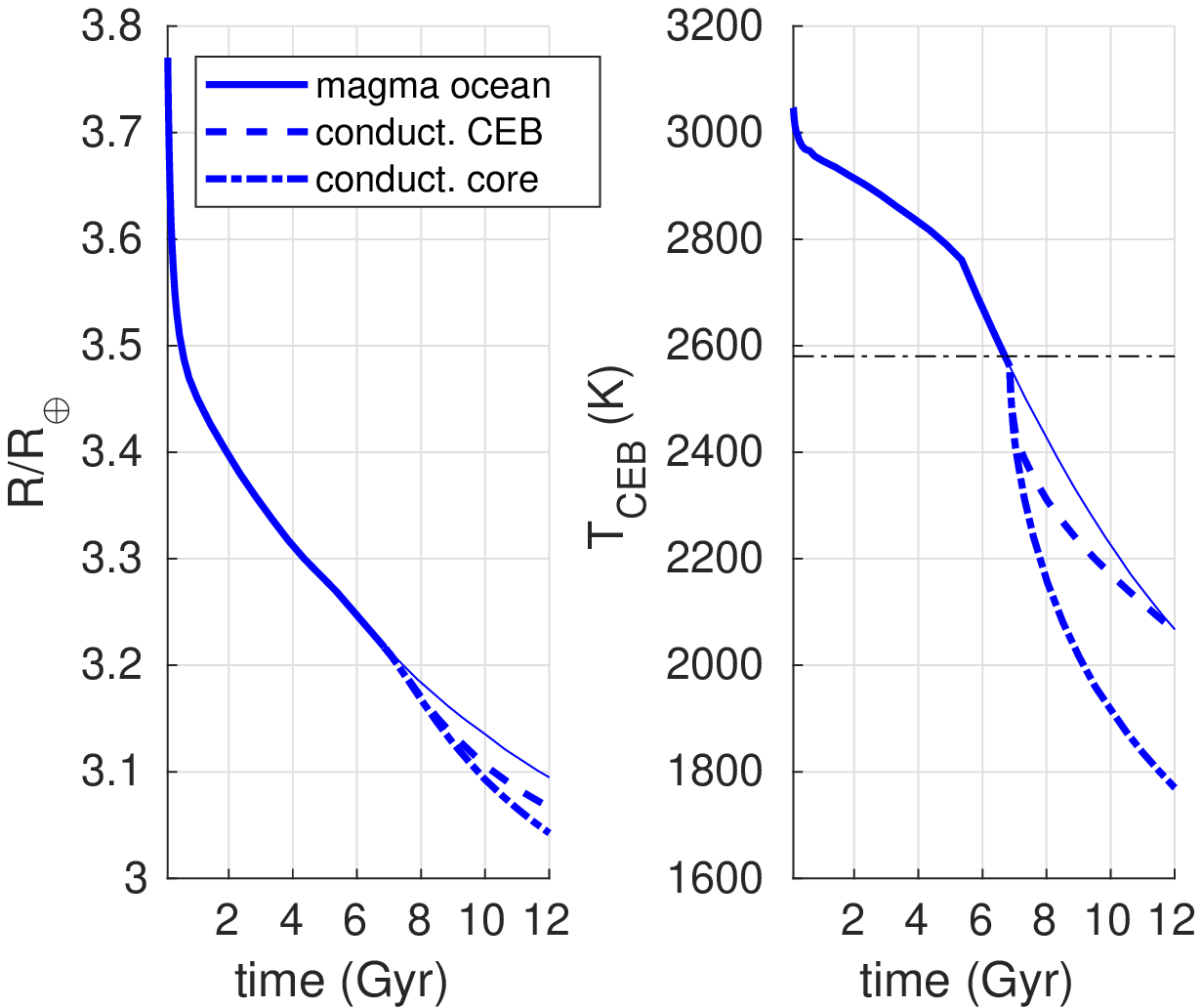}}
\centerline{\includegraphics[width=12cm]{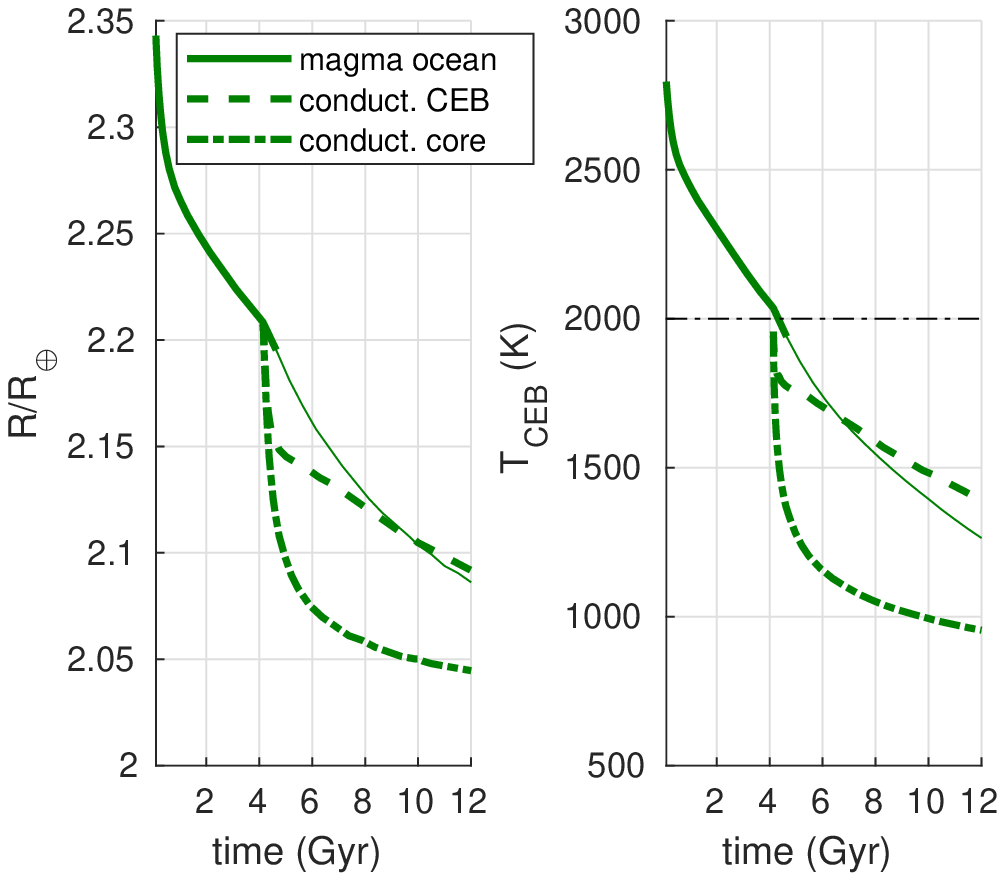}}
\caption{Radius (left) and CEB temperature (right) for a 4.5\me core with an envelope. Top panels: 10\% envelope mass (0.5\me); 
bottom panels: 0.1\% envelope mass (0.005\me). The solid curve is for efficient core cooling (magma ocean). The evolution with a conductive CEB layer (dashed) and with a conductive core (dashed-dotted) are shown after the CEB reaches the solidification temperature (horizontal dashed). }
\label{fig_TR_d}
\end{figure}

\subsection{Envelope conditions and planet mass}
Since the overall thermal evolution of the core depends also on the thermal evolution of the envelope, the thermal properties of the envelope {are expected to} change the results. 
Therefore, we calculate for enhanced (30$\times$solar) atmospheric opacity, and higher irradiation by the star (d=0.3\,AU).
In figure~\ref{fig_atms} we show the radius (left) and CEB temperature (right) in time for identical planets with different atmospheric conditions. The standard case (blue) is compared to a case with enhances opacity (dashed-dotted) and to a case with a stronger irradiation (dashed). Additional representative results appear in Table~\ref{tab_radius}.
As is shown in the figure, irradiation has a significant effect on the envelope radius\footnote{Irradiation extends the outer atmosphere. Since our evolution model is calculated for a fixed mass, we verify that the outermost layer is at p$>$10\,millibar, to avoid "fake" radii by extension of the outermost layers.}, {within the range of} previous works \citep{lopezfor14,howebur15}.
However, irradiation doesn't change the surface conditions for significant envelope masses ($>$0.01\,\me).
As the outer layers of the envelope expands by irradiation, the conditions at the bottom of the envelope remain similar to the standard case.
Thus, the location of the planet (for d$>$0.3AU) has a small effect on the core evolution in the presence of a significant envelope (photo-evaporation is not included in the model). 

Atmospheric opacity, on the other hand, changes both the radius and the core surface temperature substantially. In {the} enhanced opacity case the envelope cooling is slower. Thus, the envelope traps the heat from the core and delays the core cooling. 
As is shown in figure~\ref{fig_atms}, metallicity of 30$\times$solar keeps the core surface hotter (molten) for much longer time than in the standard case.
In the case of lower mass envelope, like for example a 4.5\me core with 0.1\% envelope, the enhanced atmospheric metallicity (30$\times$solar) delays the surface solidification from less than 0.5\,Gyr to more than 2\,Gyr.
Moreover, the effect of core thermal properties on the radius become more important for the high atmospheric opacity.
Uncertainty in radioactive elements, in the range of 0.5-2.5 Earth-ratio, results in up to 15\% radius change for planets with enhanced envelope metallicity, in comparison to only 5\% for similar planets with solar-metallicity.

\begin{figure}[ht]
\centerline{\includegraphics[width=15cm]{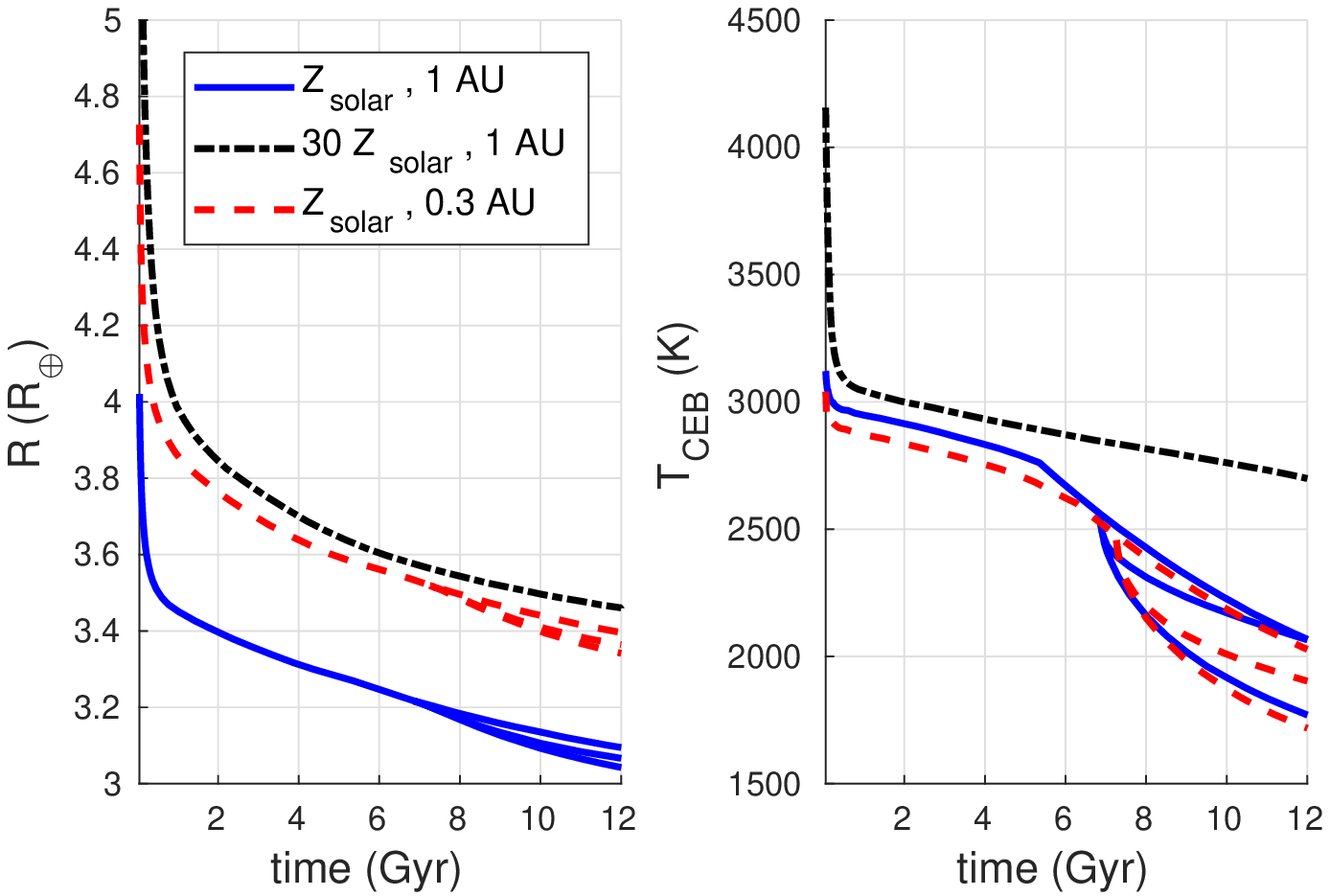}}
\caption{Radius (left) and CEB temperature (right) for 5\me planets with 0.5\me (10\%) envelope. The different curve styles are for different envelope conditions: standard case of solar metallicity opacity located at 1\,AU (solid blue), planet with enhanced atmospheric opacity of 30$\times$solar (dashed-dotted black), and planet with higher irradiation as located at 0.3\,AU (dashed red). The different curves of each color represent the different core cooling scenarios after core solidification (phase 3).}
\label{fig_atms}
\end{figure}

We also vary the mass of the core by a factor of two (2.25\me and 9$\,{\rm M}_{\oplus}$) for different envelope masses (examples in Table~\ref{tab_radius}).
We find that cores of different masses with the same envelope mass result in similar cooling rates of the core surface\footnote{It should be noted that modification of the core heat transport by the change in mass \citep[\eg][]{stamenkovic12} is not included in our simplified model.}, \ie the envelope mass, and not the envelope fraction, is the {key} parameter {driving the core thermal evolution}. The reason is that during the magma ocean phase the pressure-temperature conditions at the bottom of the envelope are determined by the adiabatic structure of the envelope. Hence, the same envelope mass leads to similar pressure-temperature conditions at the core surface.
In figure~\ref{fig_mass} we show this trend for two planets with different total masses, but the same envelope mass: the difference in magma ocean duration (CEB temperature) between a 10\me planet with 5\% envelope and a 5\me planet with 10\% envelope (same envelope mass) is smaller than the uncertainty within the core thermal properties for each planet. While the two cores feature a similar thermal evolution until solidification, the solidification point is somewhat different. The 10\me planet reaches the CEB solidification earlier than the 5\me planet, because of higher surface pressure which increases the melting (solidification) temperature.
The radii of the 5\me planet is larger, because of irradiation effect on cores with lower gravity \citep[\eg][]{lopezfor14}.
Conversely, the higher (lower) solidification temperature for more (less) massive cores, due to surface pressures, and the sensitivity to thermal effects by the higher (lower) gravity, are found to have only small effects on the radius variation by the core properties. 
Also here, high atmospheric opacity intensifies the core thermal effects. 

\begin{figure}[ht]
\centerline{\includegraphics[width=15cm]{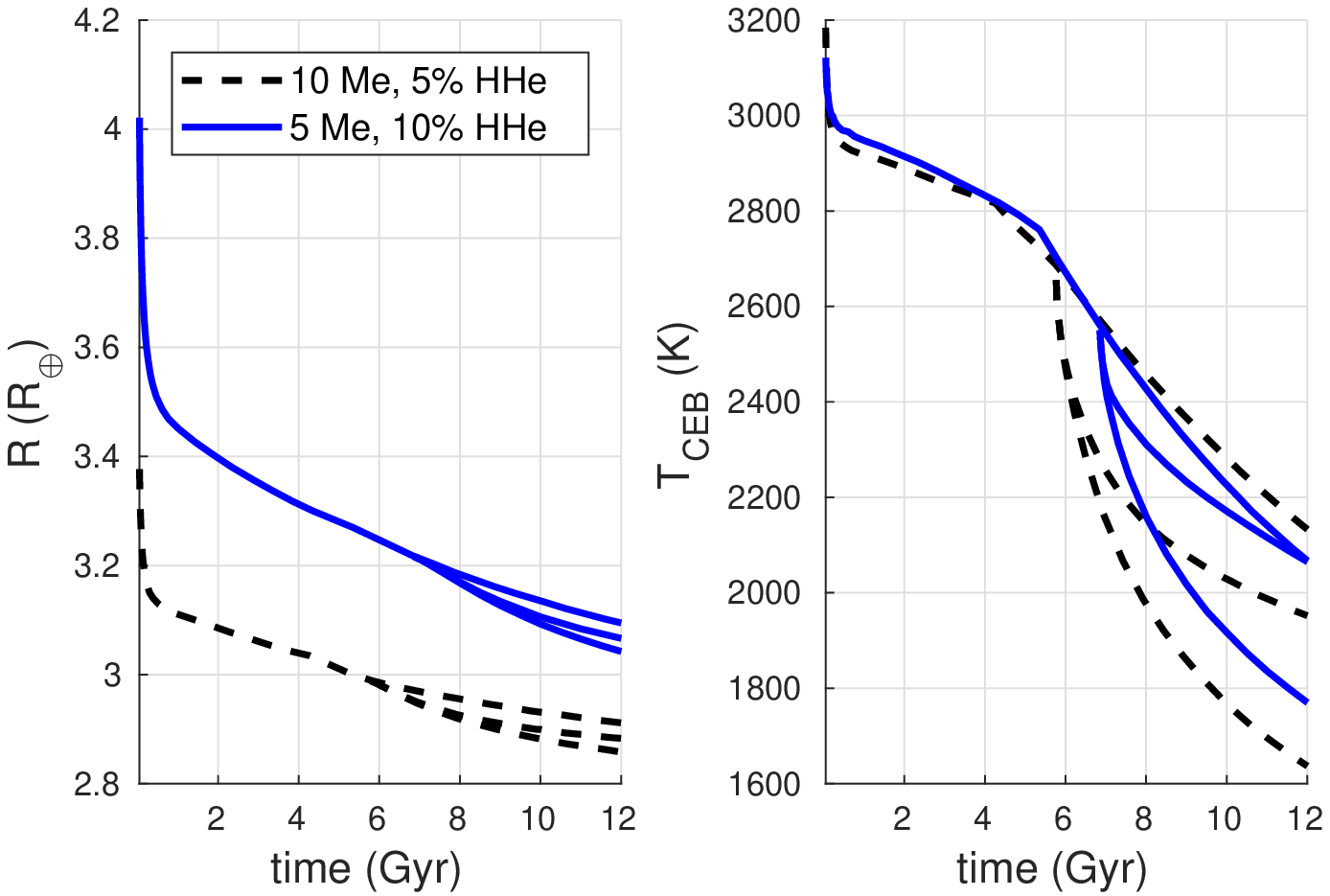}}
\caption{Radius (left) and CEB temperature (right) for 10\me (dashed black) and 5\me (solid blue) planets with the same envelope mass of 0.5\me. The different curves of each planet represent the different core cooling scenarios after solidification (phase 3). The cores feature a similar thermal evolution until solidification.}
\label{fig_mass}
\end{figure}

\begin{table*}[ht]
\centering 
\begin{tabular}{c c l c c c c} 
 \hline\hline\\[-1.5ex] 
Model &  & & t$_{\rm MO}$ (Gyr) & & radius [\re] &\\         
\hline\\[-1.5ex]
 M$_{\rm core}$& M$_{\rm env}$ & & & 1\,Gyr & 5\,Gyr & 10\,Gyr \\         
\hline\\[-1.5ex]

4.5\me & 0.5\me &  std & 6.8$^{+1.5}_{-2.3}$ & 3.45$^{+0.11}_{-0.07}$ & 3.28$^{+0.06}_{-0.08}$ & 3.14$^{+0.03}_{-0.05}$\\[+1.5ex] 
& & $\kappa_{\rm op}$ (30$\times Z_{\sun}$) & $>$13 & 4.00$^{+0.21}_{-0.12}$ & 3.71$^{+0.06}_{-0.12}$ & 3.50$^{+0.04}_{-0.04}$\\[+1.5ex] 
& & d (0.3\,AU) & 6.9$^{+2}_{-2.2}$ & 3.86$^{+0.18}_{-0.1}$ & 3.60$^{+0.08}_{-0.09}$ & 3.44$^{+0.04}_{-0.05}$ \\[+1.5ex] 

4.5\me & 0.05\me & std & 4.6$^{+1.4}_{-2.4}$ & 2.26$^{+0.05}_{-0.03}$ & 2.19$^{+0.03}_{-0.08}$ & 2.10$^{+0.02}_{-0.06}$ \\[1.5ex] 
& & $\kappa_{\rm op}$ (30$\times Z_{\sun}$)& $>$13 & 2.45$^{+0.07}_{-0.03}$ & 2.36$^{+0.07}_{-0.03}$ & 2.27$^{+0.06}_{-0.0}$\\[1.5ex] 
& & d (0.3\,AU) & 4.6$^{+1.2}_{-1.9}$ & 2.44$^{+0.09}_{-0.01}$ & 2.33$^{+0.07}_{-0.02}$ & 2.24$^{+0.04}_{-0.01}$ \\[1.5ex] 

4.5\me & 0.005\me & std & 0.2$^{+0.4}_{-0.05}$ & 1.92$^{+0.02}_{-0.07}$ & 1.88$^{+0.02}_{-0.05}$ & 1.85$^{+0.01}_{-0.03}$ \\[1.5ex] 
& & $\kappa_{\rm op}$ (30$\times Z_{\sun}$)& 2.4$^{+1.8}_{-1.1}$ & 2.04$^{+0.05}_{-0.04}$ & 1.96$^{+0.03}_{-0.03}$ & 1.93$^{+0.01}_{-0.02}$ \\[1.5ex] 
& & d (0.3\,AU) &  0.4$^{+1.2}_{-0.1}$ & 2.04$^{+0.03}_{-0.03}$ & 1.98$^{+0.02}_{-0.02}$ & 1.96$^{+0.01}_{-0.02}$ \\[1.5ex] 
\hline\\[-1.5ex]

2.25\me & 0.05\me & std & 3.5$^{+1.7}_{-1.9}$ & 2.39$^{+0.06}_{-0.07}$  & 2.20$^{+0.06}_{-0.07}$  & 2.07$^{+0.03}_{-0.04}$ \\[1.5ex] 
& & $\kappa_{\rm op}$ (30$\times Z_{\sun}$)& 11.4$^{+1.6}_{-1.4}$ &  2.89$^{+0.2}_{-0.2}$ & 2.53$^{+0.13}_{-0.06}$ & 2.41$^{+0.04}_{-0.04}$ \\[1.5ex] 
& & d (0.3\,AU) & 3.2$^{+2.1}_{-1.4}$ & 2.86$^{+0.14}_{-0.13}$ & 2.54$^{+0.13}_{-0.07}$ & 2.39$^{+0.04}_{-0.03}$ \\[1.5ex] 

2.25\me & 0.005\me & std & 0.2$^{+0.6}_{-0.2}$ & 1.82$^{+0.03}_{-0.06}$  & 1.72$^{+0.03}_{-0.04}$  & 1.68$^{+0.01}_{-0.03}$ \\[1.5ex] 
& & $\kappa_{\rm op}$ (30$\times Z_{\sun}$)& 3.5$^{+1.3}_{-1.7}$ &  2.05$^{+0.07}_{-0.11}$ & 1.87$^{+0.04}_{-0.04}$ & 1.82$^{+0.01}_{-0.03}$ \\[1.5ex] 
& & d (0.3\,AU) & 0.8$^{+0.8}_{-0.3}$ & 2.06$^{+0.04}_{-0.8}$ & 1.92$^{+0.04}_{-0.02}$ & 1.88$^{+0.01}_{-0.02}$ \\[1.5ex] 

9\me & 1\me & std & 0.1$^{+0.01}_{-0.03}$ & 3.58$^{+0.07}_{-0.03}$ & 3.44$^{+0.06}_{-0.03}$ & 3.36$^{+0.02}_{-0.02}$ \\[1.5ex] 
& & $\kappa_{\rm op}$ (30$\times Z_{\sun}$)& 3.7$^{+1.9}_{-1.6}$ & 3.96$^{+0.1}_{-0.04}$ & 3.71$^{+0.06}_{-0.04}$ & 3.6$^{+0.04}_{-0.02}$ \\[1.5ex] 
& & d (0.3\,AU) & 0.1$^{+0.03}_{-0.01}$ & 3.78$^{+0.11}_{-0.05}$ & 3.6$^{+0.06}_{-0.03}$ & 3.52$^{+0.02}_{-0.03}$ \\[1.5ex] 

9\me & 0.5\me & std & 5.3$^{+1.4}_{-2.4}$ & 3.11$^{+0.05}_{-0.04}$ & 3.02$^{+0.02}_{-0.05}$ & 2.93$^{+0.02}_{-0.03}$ \\[1.5ex] 
& & $\kappa_{\rm op}$ (30$\times Z_{\sun}$)& $>$13 & 3.38$^{+0.09}_{-0.06}$ & 3.22$^{+0.05}_{-0.05}$ & 3.12$^{+0.03}_{-0.02}$\\[1.5ex] 
& & d (0.3\,AU) & 5.7$^{+1.4}_{-2.2}$ & 3.27$^{+0.07}_{-0.07}$ & 3.13$^{+0.04}_{-0.05}$ & 3.06$^{+0.01}_{-0.03}$ \\[1.5ex] 
 \hline 
\end{tabular}
\caption{Parameter study}
\tablecomments{Planet radii and the duration of the magma ocean phase (t$_{\rm MO}$) for various planets. Results are {given} for the standard ({std}) core model (blue curve in Fig.~\ref{fig_std}-\ref{fig_TR_d}), {the high atmosphere opacity model ($\kappa_\mathrm{op}$), and the close-in model (d=0.3\,AU). Error bars indicate} the variation {of $t_\mathrm{MO}$ and radius resulting from the variation} in core properties (as in Fig.~\ref{fig_std}). For $t>t_{\rm MO}$ radius error bars include the solid state phase.\\
 }\label{tab_radius}
\end{table*}

\section{Discussion}\label{dscs}

Our results indicate that many of the observed \sn planets are in the magma ocean phase for several Gyrs. This is very different than the short magma ocean phase of planets in our solar system \citep[\eg][]{elkins12}. 
{Because of its prolonged existence, it is likely that core and envelope interact strongly; not only regarding their thermal properties (as modeled here) but also regarding their composition.}
For example, solubility of the envelope hydrogen in the silicate melt \citep{hirschmann12b} can be significant for high temperature and pressure conditions as in \sns \citep{chachan18}. 
Consequently, \sn planets with less than 1\me envelope may contain significant volatiles fraction in their cores, which affects the rock thermal and physical properties.

Conversely, the long interaction of the molten convective core with the hydrogen envelope may also enrich the envelope with metals from the molten core.
Currently, there is a lack of knowledge (experimental data as well as modelling) of rock-envelope interaction for \sn conditions.
Most of the current knowledge pertains to processes on Earth-like terrestrial planets \citep[\eg][]{abe86,hirschmann12a}.
Studies of rock-envelope interaction for \sn pressure-temperature conditions are necessary to improve our understanding of the structure and thermal properties of these objects.

As we show in this work, the envelope mass determines the state and the cooling rate of the core. 
The Kepler's data reveals a statistical dip in planetary radius between two picks of 1.3\re and 2.6\re \citep{fulton17}. This valley, which divides the close-in planets into two populations, can be a result of envelope mass loss by photoevaporation \citep{owenwu17,jinmorda18}. Envelope mass loss during the evolution shorten the magma ocean phase (as long as $t_{\rm evap}<t_{\rm MO}$) and thus affects the thermal evolution of the core.
Thus, the valley actually divides the planets also into two thermal populations, where the radius pick of 1.3\re is for bare (solid state) cores, and the 2.6\re pick is for magma ocean phase cores (M$_{\rm env}>$0.05\me).
Future studies of magma-envelope chemical interaction can provide atmospheric markers to be detected, in order to distinguish between the two populations.


This study is focused on \sn planets around sun-like star (G-type). 
\ssn planets around M-type stars, which appear to be common \citep[\eg][]{mulders15}, may have different formation environment \citep[\eg][]{kennedy08,ormel17}, and thus different core thermal contribution.
For example, different core formation timescale in a lower mass disk will change the formation energy left in the core. Moreover, low mass disk may have different composition. As a result the differentiation energy, which is derived from the iron-to-rock ratio will change.
The radioactive element abundances and the rock latent heat, which we find to contribute the most for the planet long term evolution, are derived from the rock mineralogy, which depends on metals abundances and thermal conditions of the building blocks in the disk.
As a result, the (Earth-like) values for core energy sources we used in this work (Secion~\ref{enrg_core}) will change for M-type stars.

This work consider rocky cores, without any fraction of ice. Ice-rich cores have substantial effects on both radius-mass relation and the atmospheric properties \citep{chenrog16}. However, during the solidification of the molten core water would be expelled from the core \citep[\eg][]{elkins08}. 
For the planetary and envelope mass range of \sns, water is found to stay in the envelope in vapor form; the surface pressure-temperature conditions don't allow for liquid water for the parameter range we studied here.
As we show in figure~\ref{fig_Tceb}, envelope mass larger than 0.02\me keeps the surface temperature above 1000\,K for more than 10\,Gyr.
Thus, if the core contains some fraction of ice, the envelope is presumably saturated with vapor and thus denser than the H, He envelope in our model. In this case, the increase in core radius by the icy (lighter) materials diminishes by the decrease in envelope radius by the higher mean molecular weight.
Moreover, ice-rich cores are expected to contain less radioactive elements, which are scaled with the rocky part. Therefore, the core energy source of radioactive heating is reduced.

Close-in planets usually have low envelope masses, and thus exhibits limited radius change by core thermal contribution, as is discussed in Section~\ref{rsolid}.
However, for close-in planets additional mechanisms, that are not included in our model, can affect the contribution of the core to the overall thermal evolution: 
(1) photoevaporation by the parent star. The early released core energy (formation and differentiation) extends the envelope radius and hence accelerates photoevaporation. Photoevaporation is expected to reduce the envelope mass, and therefore shorten the magma ocean phase duration. Including photoevaporation simulation \citep[\eg][]{murrayclay09} in the core-envelope evolution model is essential for modeling of close-in planets, and we would like to address it in a future work.
(2) close-in planets experience tidal forces by the parent star. Tidal heating is a continuous energy source in the core, and may be significant for super-Earth cores\footnote{corrected sentence} \citep{efroimsky12}.
(3) late planeteseimal capture can add energy during the planet evolution, as well as increase the atmospheric metallicity. This contribution is relevant in particular for planets less massive than 10\me and with envelope mass fractions less than 10\% \citep{chattchen18}.

The core model appears in this work is a first order estimate for core thermal evolution. 
In detailed (envelope-less) geophysical models \citep[\eg][]{vandenBerg19}, the transfer between the different core thermal phases is continuous, and the heat transport depends on thermal and physical properties of the core minerals. 
As we show here the core-envelope thermal effects are mutual, \ie the envelope changes the core thermal state in time.
Thus, in order to improve the existing \sn models one should link geophysical core model to envelope thermal evolution. Such a combined model is challenging, because of its self-dependent nature, but is necessary in order to better understand interiors of \sn planets.

\section{Conclusion}\label{cncld}
We have modeled the thermal evolution of \sn planets with core and hydrogen-helium envelope on one structure grid. 
Our model divides the evolution of the core into three phases: initial (formation phase), efficient cooling (magma ocean phase), and inefficient cooling (solid state phase).
We have examined the contribution of the core energy sources to the thermal evolution of the planet as a whole. In particular, we followed the mutual core-envelope thermal effects on the core solidification and the planet radius evolution.

We summarize our main conclusions below:
\begin{enumerate}
    \item Most of the observed \sn planets are in the magma ocean phase (molten surface). We find that the duration of the magma ocean phase for planets with envelope masses between 0.01\me\,--\,1\me lies between 1 and 7 Gyrs. 

    \item {Because of its efficient cooling, the magma ocean phase renders the evolution insensitive to the initial conditions. For this reason, the initial thermodynamic state of the core (heat of formation and iron differentiation) does not influence the radius evolution for more than several 10$^7$ years.}

    \item Radioactive {decay} is the most significant energy source to affect the planet radius, and the latent heat from solidification is the second. 
    In the long term, {the planet} radius variation {as a result of uncertainty in these} core energy sources is at most 15\%.

    \item After solidification of the CEB the heat flux from the core decreases further. We calculate that the variation in radius due to uncertainties regarding the post-solidification phase is no larger than 6\%.

    \item Overall, for typical model parameters, the contribution from the thermal state of the  core to the planet radius is rather limited (a few percent at most). Therefore, the inferred envelope mass from mass-radius relation is mostly proportional to the envelope (H/He) mass fraction.

\item The atmospheric opacity significantly prolongs the magma ocean phase, and amplifies the core effects on the envelope evolution.
Irradiation (for d$>$0.3AU), on the other hand, has only a minor effect on the core evolution. 
\end{enumerate}

\begin{acknowledgments}
We thank Dave Stevenson, Wim van Westrenen, and Arie van den Berg for useful comments and discussions. A.V.\ thanks Attay Kovetz for helpful ideas for code modifications. A.V.\ acknowledges support by the Amsterdam Academic Alliance (AAA) Fellowship, and by the Netherlands Origins Center.
C.W.O.\ is supported by the Netherlands Organization for Scientific Research (NWO; VIDI project 639.042.422)
\end{acknowledgments}

\begin{appendix}
\underline{A1. Rock melting curve}\\
For the calculation of rock melting curve we use solidus and liquidus pressure-temperature relation as followed:
for pressures below 10\,GPa we take melting curves for peridotite, based on \cite{deSmet99}, and consistent with melting profiles from \cite{hirschmann00} \citep[see also][]{noack17}:
\begin{equation*}
T_{\rm solid} = 1409.15\, K + 134.2\, p - 6.581\, p^2 + 0.1054\, p^3 
\end{equation*}
\begin{equation*}
T_{\rm liquid} = 2035.15\, K + 57.46\, p - 3.487\, p^2 + 0.0769\, p^3
\end{equation*}
For pressures above 10\,GPa we use melting curves for perovskite, based on \cite{stamenkov11} for the solidus. We use a fixed difference between solidus and liquidus for lack of experimental data: 
\begin{equation*}
T_{\rm solid} = 1835\, K + 36.918\, p - 0.065444\, p^2  + 0.000076686\, p^3 - 3.09272\times 10^{-8}\, p^4 
\end{equation*}
\begin{equation*}
T_{\rm liquid} = 1980\, K + 36.918\, p - 0.065444\, p^2  + 0.000076686\, p^3 - 3.09272\times 10^{-8}\, p^4
\end{equation*}

We define our melting curve at 40\% between the solidus and liquidus.
We assume that up to ~40\% melt the rock behaves like a solid, and above that as a magma ocean. 
Thus, the critical melting temperature that we use as a transition between a rock behaving as a solid and a rock behaving as a liquid is:
\begin{equation*}
T_{\rm melt} = 1659.55\, K + 103.504\, p - 5.3434\, p^2 + 0.094\, p^3
\end{equation*} 
for ${\rm p<10\,GPa}$, and 
\begin{equation*}
T_{\rm melt} = 1893\, K + 36.918\, p - 0.065444\, p^2 + 0.000076686\, p^3 - 3.09272\times 10^{-8}\, p^4
\end{equation*}
for ${\rm p>10\,GPa}$.
In figure~\ref{fig_melt} we show the resulting rock melting temperature as a function of the CEB pressure for the range of \sn planets.
\begin{figure}[ht]
\centerline{\includegraphics[width=10cm]{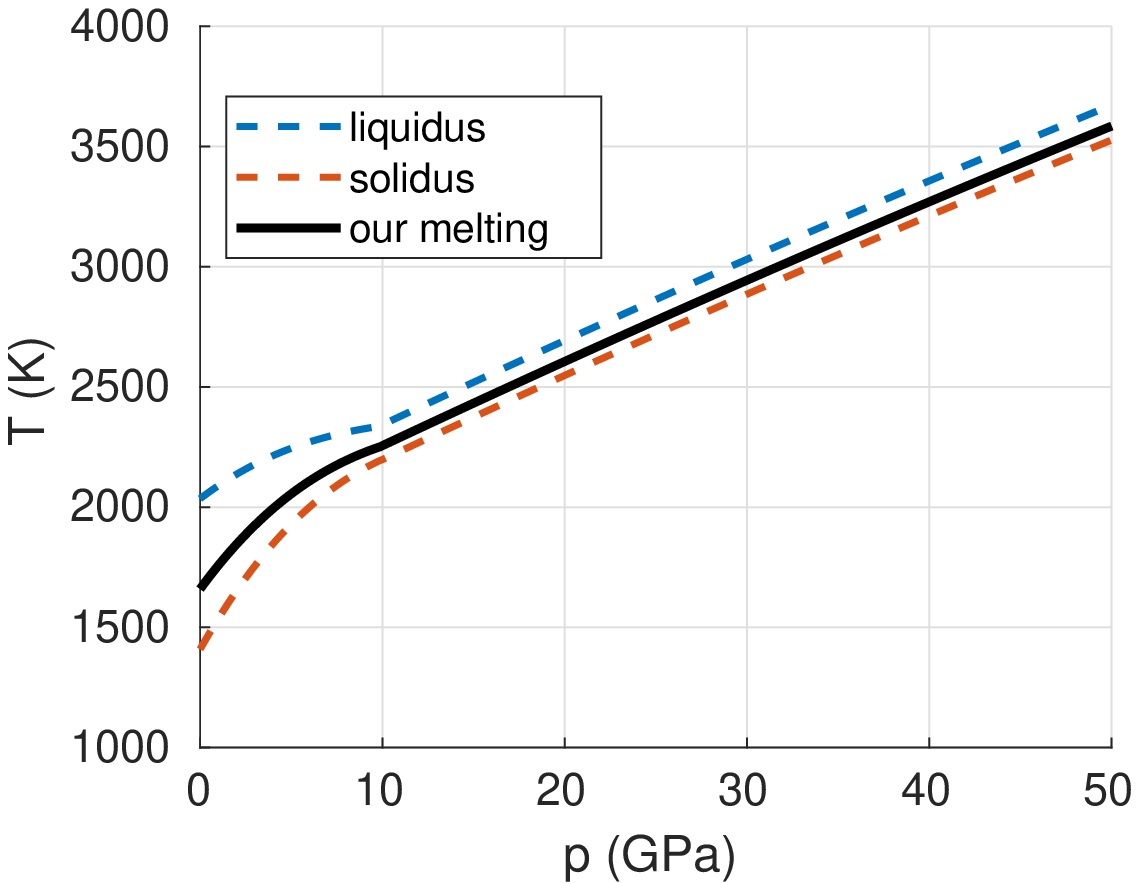}}
\caption{Rock melting curve we use in this work to determine the rocky core solidification. We define our melting curve (solid black) at 40\% between the solidus (red-dashed) and liquidus (blue-dashed). See appendix A1 for details.} 
\label{fig_melt}
\end{figure}
\\
\\

\underline{A2. Envelope mass loss}\\
As we show in Section~\ref{rslt}, the initial core energy content doesn't affect the long term radius evolution. However, this energy can bloat the envelope until it isn't bound to the planet any more \citep{ginzburg16,ginzburg17}.
Here, we test this idea by adding high fractions of the formation core binding energy to the early phase core, and following the radius of the planet in comparison to its Hill radius.
The expansion of the envelope is an outcome of the self-consistent core and envelope thermal evolution, accounting for the core heat flux and cooling, material properties (\eg tabular EOSs and opacity) and their time dependency. 
We find that {\it part} of the envelope is being lost ($\rm R_{p}>R_{Hill}$) when high fractions of formation energy is initially stored in the core. 
As the envelope expands from the core luminosity, the inner part of the envelope remains gravitationally bound to the core, and later on it cools more efficiently (than a thick envelope) and contracts. 
What fraction of the envelope is removed depends mainly on the initial core energy content, atmospheric opacity profile, and the distance from the star (Hill sphere). 
Under the conditions of our model, 
we find that envelope loss by the core energy starts when more than 30\% of the accretion energy remains in the core after its formation. For core energy content of 40\%, for example, our standard model of 4.5\me core cannot retain more than 0.1\me envelope. 
If less than 20\% of the core binding energy is left in the core after its formation, all the planetary envelopes within the range of this work survive.

\end{appendix}

\bibliographystyle{apj} 
\bibliography{allona}   

\end{document}